\documentclass[sigconf,natbib=true,anonymous=false]{acmart}
 \copyrightyear{2026}
\acmYear{2026}
\setcopyright{cc}
\setcctype{by}
\acmConference[CIKM '26] {Proceedings of the 35th ACM International Conference on Information and Knowledge Management}{November 7--11, 2026}{Rome, Italy.}
\acmBooktitle{Proceedings of the 35th ACM International Conference on Information and Knowledge Management (CIKM '26), November 7--11, 2026, Rome, Italy}
\acmISBN{979-8-4007-2539-5/2026/11} 
 \acmDOI{10.1145/3799682.3839951}
\usepackage{cleveref}
\usepackage{tcolorbox}
\usepackage{listings}
\usepackage{color,soul}
\tcbuselibrary{listings,breakable}

\newcommand{\toarrow}{\raisebox{-0ex}{\(\to\)}}
\newcommand{\zeroup}{\raisebox{-0ex}{\(^0\)}}
\newcommand{\elt}{\raisebox{-0ex}{\(\in\)}}

\newtcolorbox{quasifinbox}[1][]{
  colback=blue!3!white,
  colframe=blue!60!black,
  arc=0pt,
  boxrule=0.4pt,
  fonttitle=\bfseries\small,
  width=\columnwidth,          
  breakable=false,
  top=4pt, bottom=4pt,
  left=6pt, right=6pt,
  #1                          
}
\newtcolorbox{krullbox}{
  colback=blue!3!white,
  colframe=blue!60!black,
  arc=0pt,
  boxrule=0.4pt,
  fonttitle=\bfseries\small,
  width=\columnwidth,      
  breakable=false,         
  top=4pt, bottom=4pt,
  left=6pt, right=6pt
}
\DeclareUnicodeCharacter{2208}{\(\in\)}
\DeclareUnicodeCharacter{2070}{\(^0\)}
\DeclareUnicodeCharacter{2264}{\(\le\)}
\DeclareUnicodeCharacter{22A4}{\(\top\)}

\author{Noah Gießing}
\orcid{0009-0006-5268-2519}
\affiliation{
  \institution{FIZ Karlsruhe}
  \streetaddress{Franklinstrasse 11}
  \city{Berlin}
  \country{Germany}
  \postcode{10587}
}
\email{noah.giessing@fiz-karlsruhe.de}

\author{Ankit Satpute}
\orcid{0000-0003-3219-026X}
\affiliation{
  \institution{FIZ Karlsruhe}
  \streetaddress{Franklinstrasse 11}
  \city{Berlin}
  \country{Germany}
  \postcode{10587}
}
\email{ankit.satpute@fiz-karlsruhe.de}

\author{Hamid Rahkooy}
\affiliation{
  \institution{FIZ Karlsruhe}
  \streetaddress{Franklinstrasse 11}
  \city{Berlin}
  \country{Germany}
  \postcode{10587}
}
\email{rahkooy@gmail.com}

\author{André Greiner-Petter}
\orcid{0000-0002-5828-5497}
\affiliation{
  \institution{University of Göttingen}
  \streetaddress{Wilhelmsplatz 1}
  \city{Göttingen}
  \country{Germany}
  \postcode{37073}
}
\email{Greiner-Petter@gipplab.org}

\author{Moritz Schubotz}
\orcid{0000-0001-7141-4997}
\affiliation{
  \institution{FIZ Karlsruhe}
  \streetaddress{Franklinstrasse 11}
  \city{Berlin}
  \country{Germany}
  \postcode{10587}
}
\email{moritz.schubotz@fiz-karlsruhe.de}

\title[Formal-Informal Mathematical Alignment through and beyond Human Annotations]{Formal-Informal Mathematical Alignment through and beyond Human Annotations}
\begin{abstract}
Formalized mathematics libraries such as Lean's \texttt{mathlib} provide machine-verifiable representations of mathematical knowledge, including definitions, theorems, and their relationships.
If effectively connected to informal mathematical resources, formal mathematics could help uncover previously implicit links between mathematical concepts and improve access to existing formalizations.
However, connections between informal and formal mathematics libraries are currently sparse and unsystematic. 
In this paper, we make first strides towards better formal integration.
We construct a dataset linking informal statements from the Stacks Project to their formalizations in \texttt{mathlib} by exploiting expert annotations.
We evaluate retrieval, LLM-based reranking, and LLM-assisted verification to discover new unannotated informal-to-formal links.
To facilitate exploration and reuse of formalized knowledge, we integrated the newly discovered alignments into the MaRDI knowledge graph, enabling mathematicians to find formal counterparts to informal mathematical content.
We discuss the difficulties and challenges in using LLMs for linking formal and informal mathematical content.
All code and data necessary for reproducing this paper can be found at: \url{https://github.com/AnkitSatpute/Formal-Informal-Mathematical-Alignment}
\end{abstract}

\keywords{Mathematical information retrieval, formalized mathematics, Lean, MaRDI, LLM-as-judge}
\begin{document}

\maketitle
\section{Introduction}
\label{sec:intro}
\textit{Formal} mathematics uses proof assistants like \textit{Lean}~\cite{leanwebsite} to express definitions, theorems, and proofs in a precise, computer-verifiable language, ensuring correctness through explicit verification of every logical step.
The possibility of automated proof checking in formalized mathematics can improve mathematical workflows in a number of ways: peer-reviewing of proofs can be partially foregone (and has been, for certain exceptionally long proofs~\cite{gonthier2008formal}), immediate feedback can help train students and AI and the argumentative structure of a proof becomes clearer~\cite{Tao2025Machine}.
However, writing formal mathematics remains difficult and time-consuming, since even simple mathematical arguments often require substantial low-level detail and a steep learning curve for proof assistants~\cite{2020CicmsurvFormMath}.
To make formalization accessible, community-driven libraries such as \texttt{mathlib} for \textit{Lean} have expanded the accessibility of formal mathematics by providing reusable collections of formalized definitions and proofs~\cite{mathlib2019lean,leanwebsite}.
Usage of these libraries is essential for driving down the detail needed in writing formal proofs. 
However, the sheer volume of statements along with their technical content makes it difficult to get an overview on the available content. 
To mitigate these issues, it would be important to integrate formal libraries within informal bases of mathematical knowledge, which novice users are more proficient in interpreting.
In particular, systems that can map informal mathematical statements to their corresponding formalized versions would significantly reduce the effort required to navigate and reuse existing formal libraries.
On the other hand, machine-readability of formal libraries also makes them a highly practical resource for knowledge graphs, such as the mathematics-specific MaRDI~\cite{mardi-portal} knowledge graph, as relationships between theorems are explicit and traceable. 
In contrast, links between informal research output can be difficult to interpret or require use of hallucination-prone language models in their creation~\cite{satpute2026aspect}.

In this work, we use expert-provided connections to construct a linked informal-formal dataset between the Stacks Project~\cite{stacks-project} and \texttt{mathlib}.
We study the ability of a state-of-the-art retrieval system to recover these links automatically.
Motivated by recent progress in large language models (LLMs) for mathematical reasoning and ranking~\cite{hubert_olympiad-level_2025, dekoninck_open_2025}, we also investigate whether LLM-as-Judge-based reranking and filtering can improve retrieval quality and identify previously undiscovered candidate alignments, thereby making reliable integration of formal and informal mathematical knowledge bases.
Concretely, we make the following contributions:
\textbf{C1:} We construct a dataset connecting statements from the Stacks Project to corresponding declarations in \texttt{mathlib} using expert-curated references. 
\textbf{C2:} We evaluate LeanSearch~\cite{gao_semantic_2025} and an LLM-based reranking method on the task of retrieving formalized counterparts of informal mathematical statements.
\textbf{C3:} We use LLM-as-judge to assess whether newly discovered informal-formal alignments are mathematically plausible and discover new formal-informal pairs.
\section{Related Work}\label{sec:RelatedWork}
\par\textbf{Corpora of aligned formal-informal data.}
Typical corpora of aligned formal and informal statements are usually created by gathering statements in either formal~\cite{jiang2024multilingual, gao2025herald}, or informal~\cite{liu2026atlas} collections and translating them into the respective other modality. 
This translation occurs through manual work or by usage of LLMs. 
Manual translations are preferred for benchmarking automated theorem provers~\cite{zheng_minif2f_2022,azerbayev_proofnet_2023,tsoukalas_putnambench_2024}, while \textit{auto(in-)formalized} data is employed to power their training~\cite{xin_scaling_2025,hubert_olympiad-level_2025,xin2024advancing}. 
Datasets based on formalizing informal statements are targeted towards problem solving and do not necessarily reflect the style of formal mathematics libraries, however.
In contrast, aligned corpora from formal libraries, such as  HERALD~\cite{gao2025herald}, are used to train autoformalization models or retrieval engines.
Our contribution differs in that we link existing statements from both informal and formal libraries based on expert annotations as ground truth with the aim of closer integration between the two modalities.

\par\textbf{LLM for formalizations}
Formal mathematics has been attracted a lot of attention from the AI community in recent years, see~\cite{yang_position_2025}. The primary focus has been on developing LLM-based theorem provers, which by now leave purely symbolic methods~\cite{blanchette_hammering_2016} behind in theorem proving capability~\cite{hubert_olympiad-level_2025}\footnote{See also \url{https://leanprover.zulipchat.com/\#narrow/channel/509682-Sphere-packing-in-8-dimensions/topic/Discussion.20for.20Autoformalization/with/575354223}}. 
Aside from being used in generation of formal mathematics, LLMs have also been evaluated in judging alignment between informal and formal statements, usually to check the correctness of autoformalized statements~\cite{hubert_olympiad-level_2025}.
The LLM-as-Judge paradigm has emerged as a widely adopted approach for automated evaluation of text generation systems, leveraging the contextual reasoning capabilities of LLMs to assess relevance, quality, or correctness~\cite{dekoninck_open_2025}.
Highly relevant for mathematicians working in formal mathematics is the task of lemma retrieval: given an overabundance of lemmas in formal mathematics libraries it is crucial to identify which lemma can be applied at a given step in a proof.
The query can either be constructed from formal input (such as the proof state) \cite{yang_leandojo_2023} or as informal mathematical text \cite{gao_semantic_2025, gao2026leansearchv2globalpremise, lu_lean_2025}. 
The standard technique in the latter case is to synthesize informal analogues to \texttt{mathlib} entities (\textit{informalization})~\cite{gao2025herald} and compare their vector embeddings to that of the query.
In this work, we specifically investigate LLMs' facilities with regards to judging relationships between items in formal libraries and informal statements in comparison to expert judgments.
\section{Linked Dataset}\label{sec:data}
\subsection{Source and Target Data}
\par\textbf{\texttt{mathlib}.} is the principal archive of formal proofs in Lean.
As of February 9th 2026, \texttt{mathlib} hosts 2,175,997 lines of code in 8,241 files, containing 124,377 definitions and 257,563 theorems, making it one of the largest publicly available collections of formal mathematics. Maintenance and expansion is handled by 759 contributors. 
\par\textbf{Stacks Project.} The Stacks Project~\cite{stacks-project} is a popular open source textbook for algebraic geometry. It is organized into 20,950 subunits called \textit{tags}, including lemmas, definitions, but also entire sections and chapters. Major formalization projects in \texttt{mathlib} have used the Stacks Project as primary resource~\cite{22schemesLeanForm, 23ProjConstructLean}. 
By prepending the @[stacks] attribute above a statement\footnote{\url{https://leanprover-community.github.io/mathlib4_docs/Mathlib/Tactic/StacksAttribute.html}}, \texttt{mathlib} authors can cite source tags from the Stacks Project. 
We identified 482 links of this form, pointing to 325 tags. These links are divided across 209 mathlib files, covering about 3\% of all files.
Coverage increases for relevant subfolders, rising to 22\%, or 35\% in the areas of Algebraic Geometry, or Field Theory, respectively. We term the resulting dataset SPmathlib
As a starting point to our experiments we evaluate the ability of state-of-the-art lemma retrieval methods to recover the links from our dataset out of all statements in \texttt{mathlib}.

We take LeanSearch~\cite{gao_semantic_2025} to be a representative retrieval engine as it is one of the most popular semantic search engines for \texttt{mathlib}.\footnote{We considered LeanFinder and LeanExplore as alternatives. However, LeanFinder's API is just a wrapper around its GUI, which makes it difficult to process in an automatized fashion. LeanExplore, on the other hand, is suffering from service outages as of May 2025.} 
LeanSearch works by creating informalized versions of all statements in \texttt{mathlib}. For informalizing a statement, linked and neighboring statements are retrieved, after which an LLM is prompted with this information to generate a corresponding informal description. Given its informalized corpus and a user-generated informal query, LeanSearch uses standard Dense Passage Retrieval to obtain the closest matching informalized statement and retrieves the associated formal statement. As an addition, we evaluate a reranking scheme, by querying Qwen3.5.-35B-A3B for a reasoning-based reranking of the top ten results retrieved by LeanSearch.\\
We consider two matching criteria. Alongside the direct matching of retrieved and annotated statement (\textbf{Statement Match}), we consider a relaxed criterion (\textbf{Module Match}), which merely requires finding a statement in the same module file of \texttt{mathlib}. As mathlib modules generally contain thematically related material, such matches can be regarded as partially relevant.
\begin{table}[]\label{leansearch_mods}
\centering
\addtolength{\tabcolsep}{-0.4em}
\begin{tabular}{p{1.9cm}p{1.5cm}p{1.5cm}p{1.5cm}p{1.5cm}}
\toprule 
Method 
& R@1  & R@5  & R@10 & MRR@10 \\
\midrule
LeanSearch
&  0.36 (0.53)& 0.65 (0.70) & 0.73 (0.74) & 0.56 (0.73)\\
LeanSearch (Reranked)
& \textbf{0.46 (0.64)}& \textbf{0.69 (0.72)}& \textbf{0.72 (0.74)} & \textbf{0.67 (0.79)}\\
\bottomrule
\end{tabular}
\caption{Retrieval results of LeanSearch on SPmathlib. Bracketed scores are evaluated based on Module Match, others on Statement Match}
\label{leansearch_results}
\end{table}
In the results shown in Table \ref{leansearch_results}, we observe that LeanSearch recovers annotated statements in less than in half of cases with its highest ranked result. Reranking is particularly effective at the top instance, pushing R@1 by 28\%. However, partial relevance, measured by \textbf{Module match} shows that more than half of first retrieved instance can be considered relevant to the query.

\section{Expanding the dataset using LLM-as-Judge}
In this section, we investigate the use of LeanSearch in combination with LLM-as-Judge to uncover presently unannotated links and obtain a measure of certainty regarding whether a retrieved result presents an actual formalization. After analyzing the nature of annotated statements in SPmathlib, we define a valid formalization based on the annotation standards of \texttt{mathlib} authors. We categorize matches as \textbf{Exact} (hypotheses and conclusions match), \textbf{Equivalent} (semantically equivalent), \textbf{Generalization} (valid under more general assumptions or conclusions), or \textbf{Impl.} (implies existence of a formalization, such as groundwork for inductive types). We also account for partial matches where substatements correspond to these categories. The LLM is prompted to classify the top retrieved result into these categories, or return \textbf{No}/\textbf{Invalid} if the match is poor or the code appears inconsistent.\footnote{The categories \textbf{Impl.} and \textbf{Invalid} are mainly attempts to mitigate tendencies of the LLM to misjudge situations, in which its understanding of the code appears to be poor.} We point to the prompt in the repository for greater detail on these categories.
We employ the same Qwen model that was used as reranker to provide these judgments,
 considering it a satisfying trade-off between reasoning prowess and size.
Given the current limitations of LeanSearch in identifying formalization candidates, we cannot expect to identify all available formalizations of statements in the Stacks Project using this pipeline. 
Rather, at this stage of the project we identify conditions under which we can state with a degree of certainty that a formalization candidate should be regarded as a valid formalization of the informal source statement.

\par\textbf{Correctness of baseline evaluations.}
\begin{table}[]
\centering
\addtolength{\tabcolsep}{-0.4em}

\begin{tabular}{ccccccc}
\toprule 
& Exact (+)& General. (+)&  Equiv. (+)&  Impl. (+)&  No (-)& Inv. (-)\\
\midrule
 \#Judged & 62 & 48 & 28 & 2 & 113 & 68 \\
Acc. (\%) & 72.6 & 58.3 &  64.2 & 50.0 & 80.5 & 63.2 \\
 \midrule

 \#Judged & 62.5 & 48.5 & 28.7 & 6.1 & 114.9 & 62.5 \\
Acc. (\%) & 77.2 & 58.9 &  55.9 & 23.8 & 82.7 & 59.7 \\
\midrule

  \#Judged & 80.6 & 74.0 & 40.6 & 3.3 & 83.0 & 41.7 \\

Acc. (\%) & 76.3 & 58.2 &  69.8 & 28.5 & 69.1 & 35.6 \\
 \bottomrule
\end{tabular}
\caption{Accuracy of different LLM judgements on SPmathlib. (+)-judgements are accurate when retrieved result is a Statement Match, (-)-judgements, when it is not. First two rows: single run, no reranking. Second and third pair of rows, average accuracies of ten runs for first retrieved/first reranked candidate, respectively.}\label{table1_llm_judge}
\vspace{-25pt}
\end{table}
We evaluated LLM judgments on SPmathlib, as shown in Table \ref{table1_llm_judge}. The LLM appears to be most accurate with \textbf{Exact} and \textbf{No} judgments, likely due to their lower ambiguity, and the high prevalence of true negatives in the top results in the latter case. While standalone accuracy appears insufficient for full automation without human checks, manual review of the \textbf{Exact} false positives revealed the LLM was correct in 14 out of 16 cases. The two errors stemmed from truncated LeanSearch output; reprompting with the full code yielded correct judgment again. Thus, \textbf{Exact} judgments are highly reliable indicators of valid formalizations. Conversely, \textbf{General.} and \textbf{Equiv.} show lower correlation with ground truth, even while significantly outperforming a random oracle  (accuracy $0.35$, see \Cref{leansearch_results}).

\par\textbf{Expanding the dataset.}
The LLM demonstrated near-perfect accuracy in identifying \textbf{exact} matches and successfully uncovered several unannotated statements that align with the Stacks Project statements.
Motivated by this, we applied our method to statements from the Stacks Project that had not yet been referenced in \texttt{mathlib}. 
We queried LeanSearch with all 20,950 Stacks Project statements and filtered results with embedding distances below 0.3.
Lower distances strongly correlated with higher matching probabilities. 
This yielded 1,810 candidate statements, of which the LLM identified 228 as \textbf{exact} matches. 
To prioritize novel insights, we focused on matches from sections and chapters not present in the original 482 annotated \texttt{mathlib} statements.
Since such sections typically introduce new mathematical concepts, a formalization within them suggests that \texttt{mathlib} may already contain formalizations beyond the scope of the ground truth.
The 228 \textbf{exact} matches 108 formal statements from 67 previously unseen sections, and 12 from 8 unseen chapters. 
Investigating them, we saw that the LLM’s judgments were accurate in 103 out of 108 cases, with only 5 requiring correction. 
Excluding the 5 erroneous judgments reduced the number of unseen chapters to 5, however.
\par\textbf{Increasing certainty through majority voting.}
Although the five misclassified statements highlight automation limits, they do not undermine the LLM's overall reliability, as manual review suggests failures stem from randomness rather than systematic error: For instance, three cases are translations of scheme-theoretic results (which have been formalized in Lean), to generalizations of schemes (formalization of which has thus far not been attempted). The model rejected 300+ similar statements of the same nature. Reprompting or larger models (e.g., Qwen3.5-122B-A10B) corrected all five instances. To mitigate random errors, we propose larger models or majority voting, accepting higher computational costs for improved automation.

We evaluated majority voting on SPmathlib (\Cref{tab:judge_accs}), running queries 10 times and grouping judgments. Accuracy consistently improves for \textbf{exact} matches and rejections, though gains for \textbf{Equivalence} and \textbf{Generalization} vary across settings. We draw two key conclusions: first, Majority voting supports \textbf{exact} match automation despite higher costs and reduced candidate counts (when considering only unanimous agreement).
Secondly, High-confidence \textbf{Equivalence} and \textbf{Generalization} judgments often misalign with \texttt{mathlib} standards, necessitating human oversight.

\begin{table}[htbp]
\centering
\label{tab:judge_accs}
\addtolength{\tabcolsep}{-0.1em}

\begin{tabular}{llllll}
\toprule
\textbf{Cat.}  & $\mathbf{\ge 2}$ & $\mathbf{\ge 4}$ & $\mathbf{\ge 6}$  & $\mathbf{\ge 8}$  & $\mathbf{\ge 10}$ \\
\midrule
Exact  &  68.1 (113) & 75.9 (83) & 81.5 (65) & 82.0 (50) & 92.6 (27) \\
General  &  54.5 (101) & 58.7 (63) & 59.2 (49) & 61.3 (31) & 68.2 (22) \\
Equiv  &  58.1 (86) & 58.1 (43) & 54.2 (24) & 56.2 (16) & 33.3 (3) \\
Implies  &  22.2 (36) & 0.0 (7) & 0.0 (2) & \text{n.a.} (0) & \text{n.a.} (0) \\
No  &  71.9 (199) & 78.8 (146) & 84.9 (119) & 90.2 (92) & 97.0 (67) \\
Invalid  &  55.3 (150) & 60.7 (84) & 64.2 (53) & 65.0 (40) & 64.0 (25) \\
\midrule
Exact & 0.72 (130) & 0.77 (77) & 0.79 (68) & 0.83 (48) & 0.90 (29) \\
General & 0.56 (110) & 0.60 (75) & 0.59 (66) & 0.59 (49) & 0.59 (22) \\
Equiv & 0.69 (85) & 0.73 (52) & 0.70 (23) & 0.78 (9) & 1.00 (2) \\
Implies & 0.33 (9) & 0.00 (2) & \text{n.a.} (0) & \text{n.a.} (0) & \text{n.a.} (0)\\
No & 0.60 (127) & 0.71 (81) & 0.79 (77) & 0.83 (53) & 0.91 (23) \\
Invalid & 0.39 (80) & 0.33 (34) & 0.32 (28) & 0.27 (15) & 0.30 (10) \\
\bottomrule
\end{tabular}
\caption{Accuracy across agreement thresholds. Parentheses indicate statement counts. Top/bottom blocks represent default and reranked rankings, respectively.}

\end{table}

\section{Discussion and Conclusion}
\par\textbf{What counts as a formalization?} Why were identified formalizations unannotated in \texttt{mathlib}? Manual analysis indicates that \hyperref[krull]{omissions are often accidental}, but deliberate choices also occurred, reflecting annotation policies. Among these are a tendency towards generalizations and \hyperref[algebra-ring]{preferring statements with longer proofs over their corollaries}, but these tendencies come with exceptions. To advance integration into informal libraries we propose identifying clear rules for identifying ``canonical'' formalizations. Alternatively, we recommend annotating clusters of near-duplicates \cite{frieder_data_2024}. Both measures could help sharpen prompting and align LLM judgments with human experts in the \textbf{Generalization} and \textbf{Equivalence} categories.

\par\textbf{Conclusion and Future Work.}
In this paper we have studied expert-annotated links between formal and informal mathematics, with the aim of broadening the integration of formal mathematics in knowledge bases for informal mathematics. We have also experimented with using LLMs for retrieving aligned these annotated statements using informal queries and presented first steps towards automated discovery of unannotated links.
We included the 482 links between \texttt{mathlib} and the Stacks Project in the MaRDI knowledge graph. We additionally plan to discuss validation of the newly found links to the Stacks Project with the \texttt{mathlib} maintainers. Finally, there are also other links between \texttt{mathlib} and informal mathematics, through citations of literature and the 1000 theorems project. These links have been placed in the MaRDI knowledge graph as well.\footnote{See for example \url{https://portal.mardi4nfdi.de/wiki/Item:Q6823921}} Future work will study use and effectiveness of our LeanSearch+LLM-as-judge pipeline in those settings as well.

\section{Acknowledgements}
This work is supported by the Deutsche Forschungsgemeinschaft (DFG, German Research Foundation) – funding numbers \href{https://gepris.dfg.de/gepris/projekt/460135501}{460135501}; \href{https://gepris.dfg.de/gepris/projekt/567156310}{567156310}; \href{https://gepris.dfg.de/gepris/projekt/554559555}{554559555}; \href{https://gepris.dfg.de/gepris/projekt/437179652}{437179652}.
We utilized AI models to proofread, enhance grammar, and improve sentence clarity. 
Every sentence in the resulting text is checked by the authors, and citations are provided wherever available. 
We take full responsibility for the text in this manuscript.
\appendix

\section{Examples}
We provide examples for the three most common matching categories: \textbf{Exact}, \textbf{Generalization}, \textbf{Equivalent}, as well as partial matchings. The latter two examples also highlight cases where a retrieved candidate represents a potential formalization that has not been annotated.
\begin{krullbox}\label{fds}
\textbf{Informal Statement:} 
Let $E/F$ be an algebraic extension of fields. Let \colorbox{yellow}{$\overline{F}$} be an algebraic closure of $F$. The following are equivalent
\begin{itemize}
   \item $E$ is normal over $F$, and

   \item for every pair $\sigma , \sigma ' \in \mathop{\mathrm{Mor}}\nolimits _ F(E, \overline{F})$ we have $\sigma (E) = \sigma '(E)$.
\end{itemize}

\vspace{0.4em}
\textbf{Annotated Formalization:}
\begin{lstlisting}[escapechar=!]
@[stacks 09HQ "stronger version replacing an algebraic closure by a normal extension"]
variable {K : IntermediateField F L} {F L} !\colorbox{yellow}{[Normal F L]}!
lemma normal_iff_forall_fieldRange_eq : Normal F K ↔ ∀ σ : K →ₐ[F] L, σ.fieldRange = K :=
...
\end{lstlisting}
The annotated statement is a \textbf{Generalization} of the informal statement, see the comment attached to the ``stacks'' tag.
\end{krullbox}
\begin{quasifinbox}\label{algebra-ring}
\textbf{Informal Statement:} 
If the equivalent conditions of Lemma 10.122.2 are satisfied then we say $R \to S$ is quasi-finite at $\mathfrak q$. We say a ring map \colorbox{yellow}{$A \to B$} is quasi-finite if it is of finite type and quasi-finite at all primes of $B$.

\vspace{0.4em}
\textbf{Annotated Formalization:}
\begin{lstlisting}[escapechar=!] 
variable {R S : Type*} [CommRing R] [CommRing S] 
  !\colorbox{yellow}{[Algebra R S]}!
@[mk_iff, stacks 00PL]
class !\colorbox{red}{QuasiFinite}! : Prop where
  finite_fiber (P : Ideal R) [P.IsPrime] :
    Module.Finite P.ResidueField (P.Fiber S) := by infer_instance
\end{lstlisting}

\textbf{Candidate Formalization:}
\begin{lstlisting}[escapechar=!] 
variable {R S T : Type*} [CommRing R] [CommRing S]
def QuasiFinite {R S : Type*} [CommRing R] [CommRing S] !\colorbox{yellow}{(f : R →+* S)}! : Prop :=
  !\colorbox{red}{@Algebra.QuasiFinite}! R S _ _ !\colorbox{yellow}{f.toAlgebra}!
\end{lstlisting}
An example of the retrieved statement deriving directly from the annotated statement: The annotated statement is \textbf{Equivalent} to the informal statement, while the candidate is an \textbf{Exact} match.
\end{quasifinbox}
\begin{krullbox}\label{krull}
\textbf{Informal Statement:} 
Suppose that $R$ is a Noetherian local ring and $x\in \mathfrak m$ an element of its maximal ideal. Then \colorbox{yellow}{$\dim R \leq \dim R/xR + 1$}. \colorbox{red}{If $x$ is not contained in any of the minimal primes of $R$ then}\\\colorbox{red}{equality holds. (For example if $x$ is a nonzerodivisor.) }\newpage

\vspace{0.4em}
\textbf{Annotated Formalization:}
\begin{lstlisting}[escapechar=!]
@[stacks 0B52 "the second inequality"]
lemma _root_.ringKrullDim_quotient...
    {x : R} !\colorbox{red}{(reg : x ∈ R⁰)}! (hx : x ∈ maximalIdeal R) :
    !\colorbox{red}{ringKrullDim (R / span {x}) + 1 = ringKrullDim R }!:=
 ...
\end{lstlisting}

\textbf{Candidate Formalization:}
\begin{lstlisting}[escapechar=!]
lemma _root_.ringKrullDim_le_ringKrull...{x : R} (hx : x ∈ maximalIdeal R) :
    !\colorbox{yellow}{ringKrullDim R ≤ ringKrullDim}!
    !\colorbox{yellow}{(R / x • (⊤ : Ideal R)) + 1}! := by
 ...
  \end{lstlisting}
The annotated lemma proves the equality case of the statement, while the (unannotated) retrieved statement proves the general inequality.
\end{krullbox}
\label{sec:discussion}

\bibliographystyle{ACM-Reference-Format}
\bibliography{references} 
\end{document}